\documentclass[twocolumn,noshowpacs,preprintnumbers,amsmath,amssymb]{revtex4}
\usepackage{graphicx}
\usepackage{epsfig}
\usepackage{dcolumn}
\usepackage{bm}
\usepackage{subfigure}
\usepackage{epstopdf}
\usepackage{color}

\newcommand {\bean}  {\begin{eqnarray*}}
\newcommand {\eean}  {\end{eqnarray*}}

\def\Rset {{\rm I \kern-.2em R}} 

\newcommand {\bce}  {\begin{center}}
\newcommand {\ece}  {\end{center}}
\newcommand {\be}   {\begin{equation}}
\newcommand {\ba}   {\begin{array}}
\newcommand {\bea}  {\begin{eqnarray}}
\newcommand {\bfi}  {\begin{figure}}
\newcommand {\ee}   {\end{equation}} 
\newcommand {\ea}   {\end{array}}
\newcommand {\eea}  {\end{eqnarray}}
\newcommand {\efi}  {\end{figure}}

\newcommand {\UNIV}   {Universit\`a }

\def\Rset {{\rm I \kern-.2em R}} 
\def\mathbbH {{\rm I \kern-.2em H}} 
\def\mathbbC {{\rm I \kern-.6em C}} 

\begin{document}
\title{Intermittency and energy fluxes in the surface layer of free-surface turbulence
}
\author{Guido Troiani$^1$}
\author{Francesco Cioffi $^2$}
\author{Angelo Olivieri $^3$}
\author{Carlo Massimo Casciola$^4$}
\affiliation{$1$ Sustainable Combustion Lab. C.R. ENEA Casaccia, 
Roem, Italy.}
\affiliation{$2$ Dip. di Ingegneria Civile, Edile e  Ambientale, \UNIV di Roma
``La Sapienza'',
             via Eudossiana 18, 00184 Roma, Italy.}
\affiliation{$3$ CNR-Insean, via di Vallerano 139, 00128 Roma, Italy.}
\affiliation{$4$ Department of Mechanical and Aerospace Engineering, 
University ``La Sapienza''€˜€˜€™, via Eudossiana 18, 00184 Rome, Italy.}
\date{\today}
\pacs{Valid PACS appear here}
\maketitle

%
\section*{Abstract}
By analyzing hot-wire velocity data taken in an open channel flow,  an unambiguous definition of surface-layer thickness is here provided in terms of
the cross-over scale between backward and forward energy fluxes.  It is shown that the turbulence in the surface layer does not conform to the classical description of two-dimensional turbulence,  since the direct energy cascade persists at scales smaller than the cross-over scale, comparable with the
distance from the free-surface.

The multifractal analysis  of the one-dimensional surrogate of the turbulent kinetic energy dissipation rate
in terms of generalized dimensions and singularity spectrum indicates that intermittency is strongly depleted
in the surface layer, as shown by the singularity spectrum contracted to a single point.

The combination of intermittency indicators and energy fluxes allowed to identify the specific nature of the surface layer as alternative to classical
paradigms of three- and two-dimensional turbulence which cannot fully capture the global behavior of turbulence near a free-surface.

-------------------------------------------------------

Fluid turbulence at the free surface is crucial in many natural contexts where transport across the interface is involved. The free-surface alters the turbulent velocity fluctuations in a surface layer near the interface whose explicit determination is still debated in the literature.
A consequence is, for instance, that inertialess floaters (e.g. phytoplankton) distribute unevenly on the interface along patch- and string-like
structures \cite{lovecchio2013time,larkin2009power},
a condition that, presumably, forced the behavioral adaptation of planktonic predators in their seek for preys
\cite{seuront1999universal, schmitt2001multifractal,seuront2004random}.

On a free-surface the boundary conditions are the vanishing of tangential shear stresses and, in case of absence of waves, the annihilation of the
vertical velocity fluctuations.
This combination of constraints allow for the presence of the sole normal-vorticity component at the
interface, such that the  energy containing structures are related to vortices connected to the
free-surface \cite{troiani2004free}. Vortical structures parallel to the interface can exist below the free-surface in a range of scales
becoming  smaller and smaller as the distance from the free-surface is reduced
\cite{Kumar,Rashidi,von2011double},
resembling two-dimensional turbulence under many respects \cite{Babiano_I}.
In particular, $k^{-5/3}$ and $k^{-3}$ power laws are expected to  emerge at low and high wave numbers, respectively,
along with an inverse energy flux from mid- to low-wave  numbers \cite{lovecchio2015upscale}, \cite{Pan_ban}.

Despite the similarity with purely 2D turbulence, the free-surface layer is substantially more complex, given
the non zero two-dimensional surface divergence associated
with exchange of mass and momentum between free-surface layer and the underneath  bulk flow.
The range of validity of the two expected scaling laws depends on the distance from the free-surface.
Moreover, resolution issues and limitations on the Reynolds number make somewhat difficult to precisely identify scaling exponents due to the relatively small extension of the respective scaling ranges.
Purpose of the present Letter is to address a different class of observables,  basically intermittency indicators, in order to detect in a more clear
way the presence and nature of the free-surface layer.

It is known that the three-dimensional direct cascade of energy is highly intermittent and the moments of the coarse-grained energy dissipation rate scale with the coarse graining length  \cite{frisch1995turbulence}. In two-dimensional turbulence, instead, the inverse cascade and the dissipation field are non intermittent \cite{dubos2001intermittency,boffetta2000inverse,Paret}. In a sequence of seminal papers \cite{benzi1984multifractal,grassberger2004measuring,meneveau1991multifractal}, the turbulent kinetic energy dissipation field is regarded
as being supported on an inhomogeneous fractal characterized by a spectrum of  fractal dimensions.
This approach is exploited in this work to assess the modification induced by the free-surface on the structure of the turbulence
in an open channel flow.  Times series of streamwise velocity component are acquired by a constant temperature hot wire probe
at two different Reynolds numbers. The spectra are consistent with the expected double scaling law. However the scaling turns out to be not
sufficiently clean to confirm the theoretical interpretation. The effect of the free-surface on the turbulence statistics is instead extremely clear when
analyzed in terms of fractal properties  of the longitudinal velocity signals. The multifractal spectrum $f(\alpha)$ of the energy  dissipation field reveals  a substantial reduction of intermittency moving from the bulk to the free-surface.
The  velocity measurements are performed in an open channel at Reynolds number ${\rm Re}=33400$ where ${\rm Re}= U_{max} H/\nu$ with $U_{\max}$ the average velocity at the free-surface,  $H$ the channel depth ($H = 133 \; {\rm mm}$) and $\nu$ the kinematic viscosity. The full length of the channel is  $27$ m and it is $0.6$ m wide. The measurement station is placed at two-thirds of the length from the inlet.
The streamwise velocity component is measured with a $1$ mm wide, $4 \; {\rm \mu m}$ thick
hot wire probe, working at constant temperature with an over heat
ratio of $1.03$. Several distances from the free-surface are considered with
$2^{20}$  samples acquired at a frequency of $3000$ Hz at each position.
\begin{figure}[t!]
\begin{center}
\includegraphics[width=8.cm]{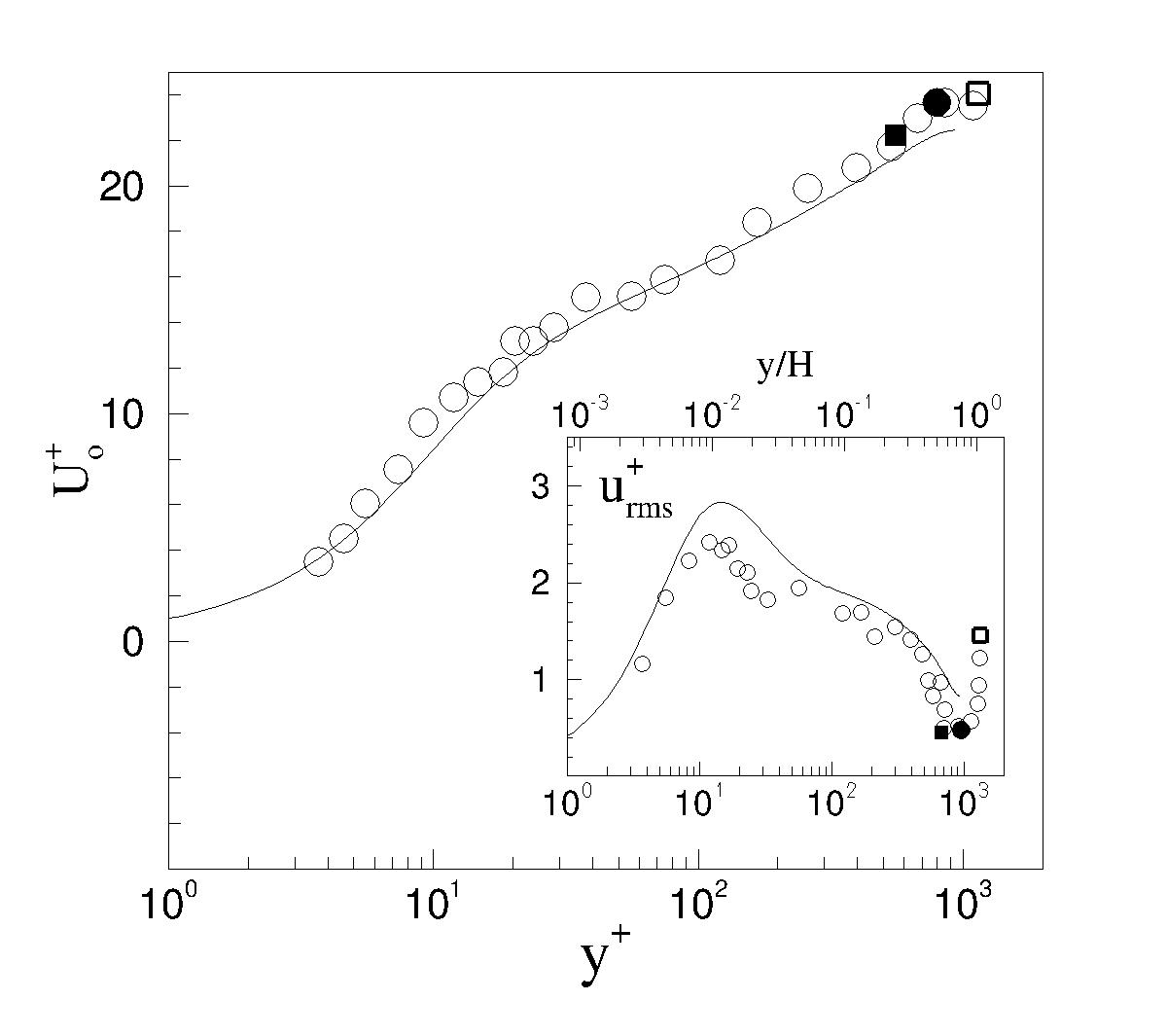}
 \caption{\small{Mean velocity profile at ${\rm Re} = 33400$, open circles. The corresponding friction Reynolds number
 is ${\rm Re}_\tau = h \sqrt{\tau_w/\rho}/\nu = 1250$. Data are compared with the (closed) channel flow
 DNS at $Re_\tau = 950$ \cite{Del_Al}, continuous line. In the inset: root mean square
  velocity fluctuations, same symbol of the mean figure.  The highlighted symbols denote the position
along the profile where the subsequent
 intermittency analysis is performed (open square, filled circle and square correspond to
$0.5\, {\rm mm}\; (y/H = .99)$,  $40 \, {\rm mm}\; (y/H = .77)$ , and $68 \, {\rm mm}\; (y/H = .49)$
below the free-surface, respectively).
  }}
 \label{Re950}
\end{center}
\end{figure}
%

Figure~\ref{Re950} shows the mean velocity profile across the channel
expressed in inner variables, $y^+ = y \sqrt{\tau_w/\rho}/\nu$,  where the shear stress $\tau_w$
at the bottom of the channel  is  evaluated {\it via} a Clauser diagram~\cite{clauser2012turbulent}.  The root mean square velocity
fluctuation, $u_{\rm rms}$, is shown in the inset.  The increase of the turbulent intensity close to the free surface is consistent
with other experimental data available in literature \cite{satoru1982turbulence}.

In the analysis described below the time signal of the streamwise velocity component $u(t)$ is converted in spatial variables $u(x)$
through the so-called Taylor hypothesis of frozen turbulence, $x = U_0 t$, where $U_0$ is the local average streamwise velocity.
Where necessary, $u' = u - U_0$ will denote the velocity fluctuation.
The reason for using Taylor hypothesis is the need for  the one-dimensional surrogate of the dissipation rate, ${\bar \epsilon} = 15 \nu
\overline {({\partial u'}/{\partial x})^2}$, where the bar denotes averaging.
%
\begin{figure}[t!]
\begin{center}
\subfigure{\includegraphics[width=8.0cm]{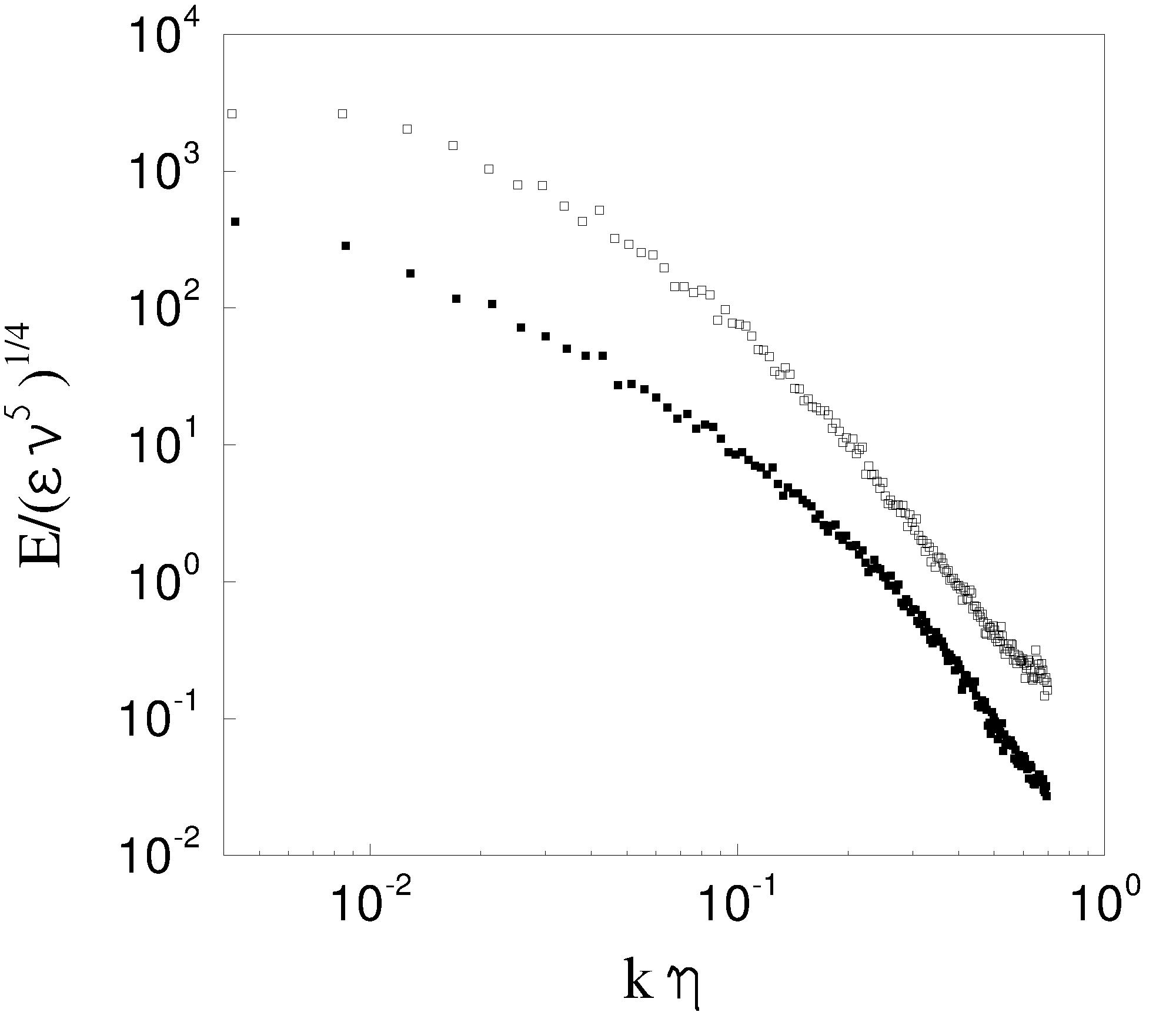}}
\caption{
\small{Velocity Spectra.  Same case as in figure~\ref{Re950}, open and filled squares denote data acquired $0.5 \, {\rm mm}$ and  $68 \, {\rm mm}$ below the free-surface, respectively. }}
 \label{S_Re950}
\end{center}
\end{figure}

Two typical spectra are shown in figure~\ref{S_Re950} corresponding to the  positions along the velocity profile highlighted by the open and filled squares  in
figure~\ref{Re950}. The spectra are plotted against the dimensionless wavenumber $k \eta$, where $\eta =
(\nu^3/{\overline \epsilon} )^{1/4}$ is the local Kolmogorov length. At large scales, low wave numbers,  the spectra approach a classical $k^{-5/3}$ scaling
range. At small scales the slope of the spectrum for the data at the free-surface is close to $-3$. It should be noted that, overall, the two spectra
do not exhibit striking differences.

\begin{figure}[t!]
\begin{center}
\includegraphics[width=8.cm]{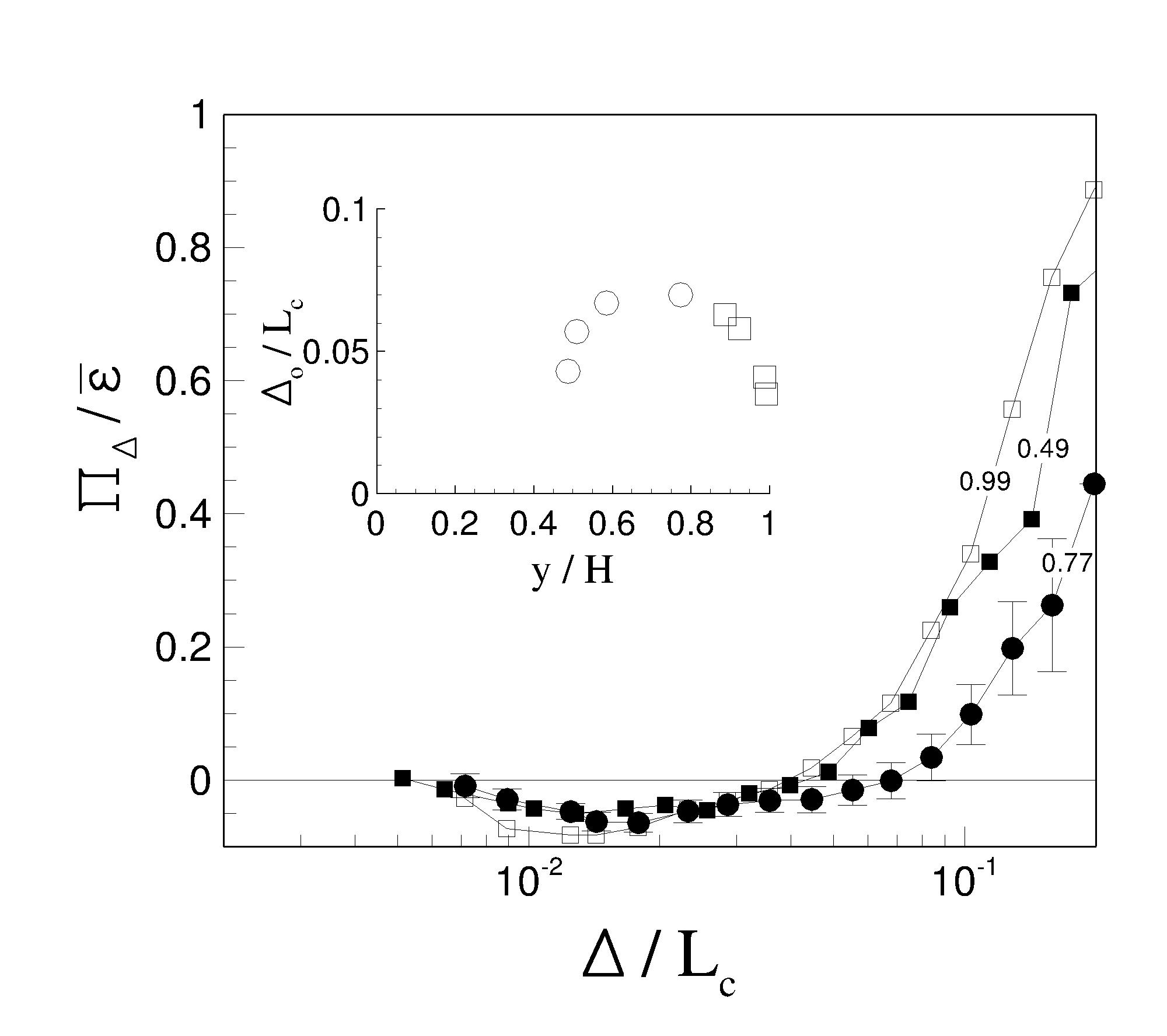}
 \caption{\small{Normalized average coarse grained energy flux $\Pi_\Delta/{\bar \epsilon}$ as a function of the coarse-graining scale $\Delta$ normalized with the correlation length $L_c$ (symbols correspond to those used in figure~\ref{Re950}.
 Inset, cross-over $\Delta_0$ between direct (at small scales) and
inverse cascade (at large scales) along the channel height.
}}
 \label{En_transf}
\end{center}
\end{figure}

It is conjectured that, in a range of scale comparable with the distance from the free-surface, the turbulent fluctuations
reproduce the behavior of two-dimensional turbulence.
If this is the case, the $k^{-5/3}$ scaling range should correspond to a backward cascade, with energy moving from
smaller to the larger scales.
The eventual forward- and backward-scattering flux of energy
is scrutinized  by applying a low-pass filter $G_{\Delta}(r) = \Delta^{-2} G(r/\Delta)$ to the velocity signal
\cite{Pan_ban,chen2006physical,lovecchio2015upscale},
like usually done in large eddy simulation. In the present case $G(r)$ is Gaussian.
Denoting by $q = u^2/2$ the (total) instantaneous kinetic energy, the corresponding coarse-grained field is
${\tilde q}(x) = \int q(y) G_{\Delta}(x-y) dy$.
The balance equation for the large-scale turbulent kinetic energy reads
\begin{eqnarray}
\partial_t {\tilde q} + \partial_j\left( {\tilde q}~ {\tilde u_j}\right)=
\partial_j\left( -2{\tilde p}~{\tilde u_j} - 2{\tilde u_i \tau_{ij}}\right)
+\frac{\displaystyle 1}{\displaystyle Re_{\tau}}\ \partial_j\left( {\tilde q}\right)
\nonumber \\
-\frac{\displaystyle 2}{\displaystyle Re_{\tau}}\ \partial_j {\tilde u_i}~\partial_j {\tilde u_i}
+ 2 \tau_{ij}\tilde{S}_{ij} ~, \qquad \qquad
\label{filt_en}
\end{eqnarray}
where $\tau_{ij} = \widetilde{u_i u_j}-\left(\tilde{u}_i \tilde{u}_j\right)$ is the
subgrid scale stress (SGS) tensor and $\tilde{S}_{ij} = \left(\partial_j \tilde{u}_i + \partial_i
\tilde{u}_j\right)/2$ the large scale rate of strain tensor.
The SGS dissipation ${\widetilde \Pi}_\Delta = \tau_{ij} \tilde{S}_{ij}$ represents the instantaneous energy
transfer across the filter scale $\Delta$.
When ${\widetilde \Pi}_{\Delta}$ is positive, the energy is transferred from smaller to larger scales,
backward energy flux. ${\widetilde \Pi}_{\Delta}$
negative would indicate a classical forward scattering of energy.
The quantity of interest here is the average SGS dissipation rate $\Pi_\Delta  = {\overline {\widetilde \Pi}}_\Delta$
shown in  figure~\ref{En_transf}  as a function of the filter width $\Delta$.
In the bulk flow negative values of $\Pi_{\Delta}$ at relatively small scales
confirm the expected direct energy transfer.  At larger scales,
energy injection by the average shear leads to cascade invertion, producing a positive energy transfer. In the bulk flow the
cross-over $\Delta_0$ between direct (at small scales) and inverse cascade (at large scales) is related to
the so-called shear scale $L_s = \sqrt{{\bar \epsilon}/S^3}$ which  characterizes the range
where energy is injected by the average shear $S = d U_0/dy$ \cite{casciola2007residual,jacob2008scaling}.
In the bulk, the cross over scale increases monotonically with increasing distance from the bottom of the channel, see the circles in the inset of the figure. At a certain critical distance the trend is
reversed and the crossover scale starts decreasing with further approach to the free surface. This is the region where the free surface directly  influences the turbulence, square symbols in the figure. In this surface layer the direct cascade occurs in a  range at small scales that becomes narrower and narrower the closer the free-surface is approached (decreasing $\Delta_0$), while an inverse cascade occurs at larger scales.
We stress that the inverse cascade in this case is not related to the injection of energy by the local shear, as it is instead the case in the bulk flow, since near the free-surface the shear vanishes. Overall, these features of the surface layer may be expected considering that the scales smaller than the distance from the free-surface are not directly influenced. On the other hand,  larger scales directly feel the constraint imposed by the interface.

The data discussed so far show that the surface definitely influences the structure of the turbulence, although  a qualitative difference does non seem to emerge in a clear way from the energy spectra. The main clue is given by the behavior of the cross-over scale $\Delta_0$ hinting at the presence of a surface layer.
As we will show below,  the multifractal description \cite{Frisch1985} turns out to be much more sensitive to the near free-surface dynamics.

The multifractal formalism is used to describe the statistical properties of the 1D turbulent kinetic energy dissipation rate estimator $\epsilon(x)$  in terms of its singularity spectrum. Since $\epsilon$ is positive definite it can be interpreted as a measure.  We cover  the support of the dissipation rate with $N$ line segments $B_i(r)$ of length $r$ and define  $P_i (r) = C \int_{B_i(r)} \epsilon(x)  dx$ the measure of the $i$th segment, where $C$ is a normalization constant.
A range of scales is identified where scaling laws of the form $P_i (r) \propto {\tilde r}^{\alpha}$ are observed, with $\alpha_{min} \le \alpha \le \alpha_{max}$ and the tilde denotes normalization with the local correlation length.
In principle, the spectrum of fractal dimensions $f(\alpha)$  can be extracted by counting the number of segments $dN(\alpha)$ where the exponent is in the range $\alpha$ and $\alpha + d \alpha$,  $dN(\alpha) \propto \rho (\alpha) {\tilde r}^{-f(\alpha)} d\alpha$, being $\rho (\alpha)$ a smooth function independent of ${\tilde r}$.

In practice, the multifractal spectrum is more conveniently evaluated starting from the  the generalized dimensions $D_q$ defined as the scaling
exponents of the $q$th moments of the measure $P_i (r)$,
$M_q(r) = \sum_i P_i^q(r) =  \left< P^{q-1}(r)\right>\propto {\tilde r}^{\left(q-1\right) D_q}$,
with $\left< \cdot \right>$ the ensemble average.
They characterize the unevenness of the measure:
moments of positive order ($q > 0$) are dominated by regions where the dissipation is concentrated,
the more so, the larger is $q$. Moments with negative order put instead more weight on low-dissipation regions.
In particular, $D_{q=0}$ is the dimension of the support, i.e. the Hausdorff or fractal dimension.

The singularity spectrum $f(\alpha)$ can be retrieved from the generalized dimensions $D_q$.
By rewriting the moments as
\begin{equation}
M_q(r) \propto \int_{\alpha_{min}}^{\alpha_{max}} d\alpha \rho (\alpha)
r^{\alpha q - f(\alpha)},
\label{sing_sp}
\end{equation}
the integral  can be evaluated using a saddle point approximation. The leading term is identified from
$d\left[q \alpha - f(\alpha)  \right]/d\alpha = 0$.
Given that $d^2\left[q \alpha - f(\alpha)  \right]/d\alpha^2 > 0$, i.e. $f''\left[\alpha(q)\right] < 0$, it follows
$f'\left[\alpha(q)\right] = q$.  Hence $M_q(r) \propto r^{q \alpha(q) - f(\alpha(q))}$, that is $\tau(q) = (q-1) D_q =q \alpha(q) - f\left[(\alpha(q)\right]$. In other words,  $\tau(q)$ and $f(\alpha)$ are the Legendre transform one of the other. Inverting the Legendre transform, we have
$f(\alpha) =  \alpha q(\alpha) - \tau\left[q( \alpha)\right]$, where $q\left(\alpha \right)$ is the solution of $d\tau(q)/dq = \alpha$.

\begin{figure}[t!]
\begin{center}
\includegraphics[width=8.cm]{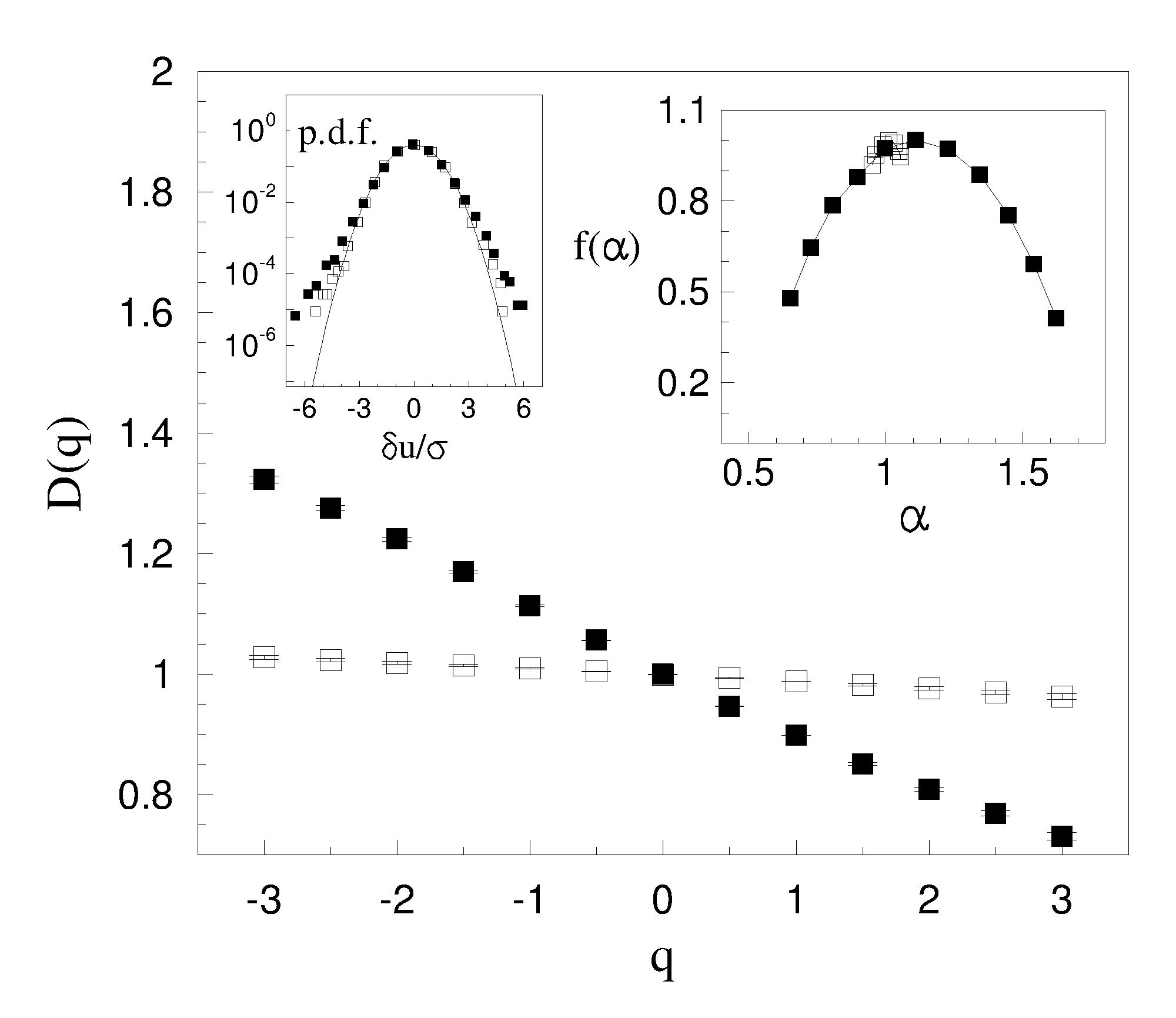}
 \caption{\small{Generalized dimension $D(q)$ {\it vs} $q$.
Left inset: pdf of the longitudinal velocity increments $\delta u/\sigma$ at different scale separations $\delta$.
Symbols: open squares, $\delta = 0.07~k \eta$, filled gradients,
$\delta = 0.5~k \eta$.
$\sigma$:  root mean square of the velocity increments.
Thin black line, reference Gaussian distribution.
Right inset: multifractal spectrum obtained by Legendre transform of $D(q)$.
Same symbols of figure \ref{Re950}.
  }}
 \label{gen_dim}
\end{center}
\end{figure}

Figure~\ref{gen_dim} illustrates the main result of the paper, showing the different moments in the range $-3 \le q \le 3$ where good statistical convergence is achieved, see Supplemental Information \cite{SI}. Two data sets are displayed, one relative to the bulk and the other to the surface layer. In both cases the data are analyzed at separations corresponding to the putative inertial range.

The data in the bulk flow, filled symbols, show a decreasing trend for the generalized dimensions as the order of the moment increases. On the contrary, for data in the surface layer, empty symbols, the generalized dimensions
are constant at increasing $q$.
In more details,  the energy dissipation rate in the  bulk flow presents singularity indices $\alpha$
ranging from $\alpha_{min}=0.64$ to $\alpha_{max}=1.54$, see the filled symbols in the inset on the right of
 figure~\ref{gen_dim}, typical of  the intermittency found in other three-dimensional configurations such as wakes
behind obstacles and atmospheric turbulence \cite{meneveau1987multifractal}.
As we move the measurement probe towards the free surface, intermittency is
considerably reduced and in the limit
where $D(q)= {\rm constant}~\forall q$ the singularity
spectrum is reduced  to a point, i.e., the system becomes monofractal, empty symbols in the
 inset  of figure~\ref{gen_dim}. As a consequence the dissipation field does not exhibit  intermittency in the surface layer.

The inset on the left of figure~\ref{gen_dim} shows the pdf of the velocity increments in log-lin coordinates to
highlight the tails of the distributions.
Reduction of intermittency, described  in terms of deviation from gaussianity, is apparent moving from the bulk (filled symbols) to the surface
layer (open symbols).

In conclusion,
the experimental result we have discussed  confirm that the free-surface deeply modifies the structure of the turbulence, in a layer close to the interface.
The inter-scale energy flux  highlights regions of forward and backward cascade both in the bulk and in the near-surface region. The backward flux close to the free-surface is however of a different nature. In particular, it is not sustained by the mean shear, as it happens in the bulk region.
We  have shown that the surface layer can be unambiguously identified by the behavior of the cross-over scale between backward and forward cascade.
At the distance from the free-surface corresponding to the thickness of the surface layer the cross-over scale present a well defined maximum.

The multifractal analysis of the dissipation field in terms of  generalized dimensions and singularity spectrum provided evidence of a strong
reduction of intermittency in the energy dissipation rate at the free-surface: while the singularity spectrum in the bulk reproduces the well known
features observed in classical three-dimensional turbulence,  the spectrum at the free surface is contracted to a single point. As a consequence the multifractal nature of the dissipation field is lost and the signal becomes non-intermittent. This conclusion is confirmed by the near-Gaussian behavior  of
velocity increments.

The free-surface layer shares several characteristics with two-dimensional turbulence: i) a range of inverse cascade at larger scales; ii) a concurrent reduction of intermittency in the dissipation field. There is however a significant difference, namely a direct energy cascade  range at small scales not found in purely two-dimensional turbulence.
This direct energy cascade is observed at longitudinal scales smaller than the distance from the free surface.

\begin{thebibliography}{28}
\expandafter\ifx\csname natexlab\endcsname\relax\def\natexlab#1{#1}\fi
\expandafter\ifx\csname bibnamefont\endcsname\relax
  \def\bibnamefont#1{#1}\fi
\expandafter\ifx\csname bibfnamefont\endcsname\relax
  \def\bibfnamefont#1{#1}\fi
\expandafter\ifx\csname citenamefont\endcsname\relax
  \def\citenamefont#1{#1}\fi
\expandafter\ifx\csname url\endcsname\relax
  \def\url#1{\texttt{#1}}\fi
\expandafter\ifx\csname urlprefix\endcsname\relax\def\urlprefix{URL }\fi
\providecommand{\bibinfo}[2]{#2}
\providecommand{\eprint}[2][]{\url{#2}}

\bibitem[{\citenamefont{Lovecchio et~al.}(2013)\citenamefont{Lovecchio,
  Marchioli, and Soldati}}]{lovecchio2013time}
\bibinfo{author}{\bibfnamefont{S.}~\bibnamefont{Lovecchio}},
  \bibinfo{author}{\bibfnamefont{C.}~\bibnamefont{Marchioli}},
  \bibnamefont{and} \bibinfo{author}{\bibfnamefont{A.}~\bibnamefont{Soldati}},
  \bibinfo{journal}{Physical Review E} \textbf{\bibinfo{volume}{88}},
  \bibinfo{pages}{033003} (\bibinfo{year}{2013}).

\bibitem[{\citenamefont{Larkin et~al.}(2009)\citenamefont{Larkin, Bandi, Pumir,
  and Goldburg}}]{larkin2009power}
\bibinfo{author}{\bibfnamefont{J.}~\bibnamefont{Larkin}},
  \bibinfo{author}{\bibfnamefont{M.}~\bibnamefont{Bandi}},
  \bibinfo{author}{\bibfnamefont{A.}~\bibnamefont{Pumir}}, \bibnamefont{and}
  \bibinfo{author}{\bibfnamefont{W.~I.} \bibnamefont{Goldburg}},
  \bibinfo{journal}{Physical Review E} \textbf{\bibinfo{volume}{80}},
  \bibinfo{pages}{066301} (\bibinfo{year}{2009}).

\bibitem[{\citenamefont{Seuront et~al.}(1999)\citenamefont{Seuront, Schmitt,
  Lagadeuc, Schertzer, and Lovejoy}}]{seuront1999universal}
\bibinfo{author}{\bibfnamefont{L.}~\bibnamefont{Seuront}},
  \bibinfo{author}{\bibfnamefont{F.}~\bibnamefont{Schmitt}},
  \bibinfo{author}{\bibfnamefont{Y.}~\bibnamefont{Lagadeuc}},
  \bibinfo{author}{\bibfnamefont{D.}~\bibnamefont{Schertzer}},
  \bibnamefont{and} \bibinfo{author}{\bibfnamefont{S.}~\bibnamefont{Lovejoy}},
  \bibinfo{journal}{Journal of Plankton Research} \textbf{\bibinfo{volume}{21}}
  (\bibinfo{year}{1999}).

\bibitem[{\citenamefont{Schmitt and Seuront}(2001)}]{schmitt2001multifractal}
\bibinfo{author}{\bibfnamefont{F.~G.} \bibnamefont{Schmitt}} \bibnamefont{and}
  \bibinfo{author}{\bibfnamefont{L.}~\bibnamefont{Seuront}},
  \bibinfo{journal}{Physica A: Statistical Mechanics and its Applications}
  \textbf{\bibinfo{volume}{301}}, \bibinfo{pages}{375} (\bibinfo{year}{2001}).

\bibitem[{\citenamefont{Seuront et~al.}(2004)\citenamefont{Seuront, Schmitt,
  Brewer, Strickler, and Souissi}}]{seuront2004random}
\bibinfo{author}{\bibfnamefont{L.}~\bibnamefont{Seuront}},
  \bibinfo{author}{\bibfnamefont{F.~G.} \bibnamefont{Schmitt}},
  \bibinfo{author}{\bibfnamefont{M.~C.} \bibnamefont{Brewer}},
  \bibinfo{author}{\bibfnamefont{J.~R.} \bibnamefont{Strickler}},
  \bibnamefont{and} \bibinfo{author}{\bibfnamefont{S.}~\bibnamefont{Souissi}},
  \bibinfo{journal}{Zool. Stud} \textbf{\bibinfo{volume}{43}},
  \bibinfo{pages}{498} (\bibinfo{year}{2004}).

\bibitem[{\citenamefont{Troiani et~al.}(2004)\citenamefont{Troiani, Cioffi, and
  Casciola}}]{troiani2004free}
\bibinfo{author}{\bibfnamefont{G.}~\bibnamefont{Troiani}},
  \bibinfo{author}{\bibfnamefont{F.}~\bibnamefont{Cioffi}}, \bibnamefont{and}
  \bibinfo{author}{\bibfnamefont{C.~M.} \bibnamefont{Casciola}},
  \bibinfo{journal}{Journal of Hydraulic Engineering}
  \textbf{\bibinfo{volume}{130}}, \bibinfo{pages}{313} (\bibinfo{year}{2004}).

\bibitem[{\citenamefont{Kumar et~al.}(1998)\citenamefont{Kumar, Gupta, and
  Banerjee}}]{Kumar}
\bibinfo{author}{\bibfnamefont{S.}~\bibnamefont{Kumar}},
  \bibinfo{author}{\bibfnamefont{R.}~\bibnamefont{Gupta}}, \bibnamefont{and}
  \bibinfo{author}{\bibfnamefont{S.}~\bibnamefont{Banerjee}},
  \bibinfo{journal}{Physics of Fluids} \textbf{\bibinfo{volume}{10}},
  \bibinfo{pages}{437} (\bibinfo{year}{1998}).

\bibitem[{\citenamefont{Rashidi and Banerjee}(1988)}]{Rashidi}
\bibinfo{author}{\bibfnamefont{M.}~\bibnamefont{Rashidi}} \bibnamefont{and}
  \bibinfo{author}{\bibfnamefont{S.}~\bibnamefont{Banerjee}},
  \bibinfo{journal}{Physics of Fluids (1958-1988)}
  \textbf{\bibinfo{volume}{31}}, \bibinfo{pages}{2491} (\bibinfo{year}{1988}).

\bibitem[{\citenamefont{Von~Kameke et~al.}(2011)\citenamefont{Von~Kameke, Huhn,
  Fern{\'a}ndez-Garc{\'\i}a, Munuzuri, and
  P{\'e}rez-Mu{\~n}uzuri}}]{von2011double}
\bibinfo{author}{\bibfnamefont{A.}~\bibnamefont{Von~Kameke}},
  \bibinfo{author}{\bibfnamefont{F.}~\bibnamefont{Huhn}},
  \bibinfo{author}{\bibfnamefont{G.}~\bibnamefont{Fern{\'a}ndez-Garc{\'\i}a}},
  \bibinfo{author}{\bibfnamefont{A.}~\bibnamefont{Munuzuri}}, \bibnamefont{and}
  \bibinfo{author}{\bibfnamefont{V.}~\bibnamefont{P{\'e}rez-Mu{\~n}uzuri}},
  \bibinfo{journal}{Physical review letters} \textbf{\bibinfo{volume}{107}},
  \bibinfo{pages}{074502} (\bibinfo{year}{2011}).

\bibitem[{\citenamefont{Babiano et~al.}(1995)\citenamefont{Babiano, Dubrulle,
  and Frick}}]{Babiano_I}
\bibinfo{author}{\bibfnamefont{A.}~\bibnamefont{Babiano}},
  \bibinfo{author}{\bibfnamefont{B.}~\bibnamefont{Dubrulle}}, \bibnamefont{and}
  \bibinfo{author}{\bibfnamefont{P.}~\bibnamefont{Frick}},
  \bibinfo{journal}{Physical Review E} \textbf{\bibinfo{volume}{52}},
  \bibinfo{pages}{3719} (\bibinfo{year}{1995}).

\bibitem[{\citenamefont{Lovecchio et~al.}(2015)\citenamefont{Lovecchio, Zonta,
  and Soldati}}]{lovecchio2015upscale}
\bibinfo{author}{\bibfnamefont{S.}~\bibnamefont{Lovecchio}},
  \bibinfo{author}{\bibfnamefont{F.}~\bibnamefont{Zonta}}, \bibnamefont{and}
  \bibinfo{author}{\bibfnamefont{A.}~\bibnamefont{Soldati}},
  \bibinfo{journal}{Physical Review E} \textbf{\bibinfo{volume}{91}},
  \bibinfo{pages}{033010} (\bibinfo{year}{2015}).

\bibitem[{\citenamefont{Pan and Banerjee}(1995)}]{Pan_ban}
\bibinfo{author}{\bibfnamefont{Y.}~\bibnamefont{Pan}} \bibnamefont{and}
  \bibinfo{author}{\bibfnamefont{S.}~\bibnamefont{Banerjee}},
  \bibinfo{journal}{Physics of Fluids (1994-present)}
  \textbf{\bibinfo{volume}{7}}, \bibinfo{pages}{1649} (\bibinfo{year}{1995}).

\bibitem[{\citenamefont{Frisch}(1995)}]{frisch1995turbulence}
\bibinfo{author}{\bibfnamefont{U.}~\bibnamefont{Frisch}},
  \emph{\bibinfo{title}{Turbulence: the legacy of AN Kolmogorov}}
  (\bibinfo{publisher}{Cambridge university press}, \bibinfo{year}{1995}).

\bibitem[{\citenamefont{Dubos et~al.}(2001)\citenamefont{Dubos, Babiano, Paret,
  and Tabeling}}]{dubos2001intermittency}
\bibinfo{author}{\bibfnamefont{T.}~\bibnamefont{Dubos}},
  \bibinfo{author}{\bibfnamefont{A.}~\bibnamefont{Babiano}},
  \bibinfo{author}{\bibfnamefont{J.}~\bibnamefont{Paret}}, \bibnamefont{and}
  \bibinfo{author}{\bibfnamefont{P.}~\bibnamefont{Tabeling}},
  \bibinfo{journal}{Physical Review E} \textbf{\bibinfo{volume}{64}},
  \bibinfo{pages}{036302} (\bibinfo{year}{2001}).

\bibitem[{\citenamefont{Boffetta et~al.}(2000)\citenamefont{Boffetta, Celani,
  and Vergassola}}]{boffetta2000inverse}
\bibinfo{author}{\bibfnamefont{G.}~\bibnamefont{Boffetta}},
  \bibinfo{author}{\bibfnamefont{A.}~\bibnamefont{Celani}}, \bibnamefont{and}
  \bibinfo{author}{\bibfnamefont{M.}~\bibnamefont{Vergassola}},
  \bibinfo{journal}{Physical Review E} \textbf{\bibinfo{volume}{61}},
  \bibinfo{pages}{R29} (\bibinfo{year}{2000}).

\bibitem[{\citenamefont{Paret and Tabeling}(1998)}]{Paret}
\bibinfo{author}{\bibfnamefont{J.}~\bibnamefont{Paret}} \bibnamefont{and}
  \bibinfo{author}{\bibfnamefont{P.}~\bibnamefont{Tabeling}},
  \bibinfo{journal}{Physics of Fluids (1994-present)}
  \textbf{\bibinfo{volume}{10}}, \bibinfo{pages}{3126} (\bibinfo{year}{1998}).

\bibitem[{\citenamefont{Benzi et~al.}(1984)\citenamefont{Benzi, Paladin,
  Parisi, and Vulpiani}}]{benzi1984multifractal}
\bibinfo{author}{\bibfnamefont{R.}~\bibnamefont{Benzi}},
  \bibinfo{author}{\bibfnamefont{G.}~\bibnamefont{Paladin}},
  \bibinfo{author}{\bibfnamefont{G.}~\bibnamefont{Parisi}}, \bibnamefont{and}
  \bibinfo{author}{\bibfnamefont{A.}~\bibnamefont{Vulpiani}},
  \bibinfo{journal}{Journal of Physics A: Mathematical and General}
  \textbf{\bibinfo{volume}{17}}, \bibinfo{pages}{3521} (\bibinfo{year}{1984}).

\bibitem[{\citenamefont{Grassberger and
  Procaccia}(2004)}]{grassberger2004measuring}
\bibinfo{author}{\bibfnamefont{P.}~\bibnamefont{Grassberger}} \bibnamefont{and}
  \bibinfo{author}{\bibfnamefont{I.}~\bibnamefont{Procaccia}}, in
  \emph{\bibinfo{booktitle}{The Theory of Chaotic Attractors}}
  (\bibinfo{publisher}{Springer}, \bibinfo{year}{2004}), pp.
  \bibinfo{pages}{170--189}.

\bibitem[{\citenamefont{Meneveau and
  Sreenivasan}(1991)}]{meneveau1991multifractal}
\bibinfo{author}{\bibfnamefont{C.}~\bibnamefont{Meneveau}} \bibnamefont{and}
  \bibinfo{author}{\bibfnamefont{K.~R.} \bibnamefont{Sreenivasan}},
  \bibinfo{journal}{Journal of Fluid Mechanics} \textbf{\bibinfo{volume}{224}},
  \bibinfo{pages}{180} (\bibinfo{year}{1991}).

\bibitem[{\citenamefont{Del~Alamo et~al.}(2004)\citenamefont{Del~Alamo,
  Jim{\'e}nez, Zandonade, and Moser}}]{Del_Al}
\bibinfo{author}{\bibfnamefont{J.~C.} \bibnamefont{Del~Alamo}},
  \bibinfo{author}{\bibfnamefont{J.}~\bibnamefont{Jim{\'e}nez}},
  \bibinfo{author}{\bibfnamefont{P.}~\bibnamefont{Zandonade}},
  \bibnamefont{and} \bibinfo{author}{\bibfnamefont{R.~D.} \bibnamefont{Moser}},
  \bibinfo{journal}{Journal of Fluid Mechanics} \textbf{\bibinfo{volume}{500}},
  \bibinfo{pages}{135} (\bibinfo{year}{2004}).

\bibitem[{\citenamefont{Clauser}(2012)}]{clauser2012turbulent}
\bibinfo{author}{\bibfnamefont{F.~H.} \bibnamefont{Clauser}},
  \bibinfo{journal}{Journal of the Aeronautical Sciences}
  (\bibinfo{year}{2012}).

\bibitem[{\citenamefont{Satoru et~al.}(1982)\citenamefont{Satoru, Hiromasa,
  Fumimaru, and Tokuro}}]{satoru1982turbulence}
\bibinfo{author}{\bibfnamefont{K.}~\bibnamefont{Satoru}},
  \bibinfo{author}{\bibfnamefont{U.}~\bibnamefont{Hiromasa}},
  \bibinfo{author}{\bibfnamefont{O.}~\bibnamefont{Fumimaru}}, \bibnamefont{and}
  \bibinfo{author}{\bibfnamefont{M.}~\bibnamefont{Tokuro}},
  \bibinfo{journal}{International Journal of Heat and Mass Transfer}
  \textbf{\bibinfo{volume}{25}}, \bibinfo{pages}{513} (\bibinfo{year}{1982}).

\bibitem[{\citenamefont{Chen et~al.}(2006)\citenamefont{Chen, Ecke, Eyink,
  Rivera, Wan, and Xiao}}]{chen2006physical}
\bibinfo{author}{\bibfnamefont{S.}~\bibnamefont{Chen}},
  \bibinfo{author}{\bibfnamefont{R.~E.} \bibnamefont{Ecke}},
  \bibinfo{author}{\bibfnamefont{G.~L.} \bibnamefont{Eyink}},
  \bibinfo{author}{\bibfnamefont{M.}~\bibnamefont{Rivera}},
  \bibinfo{author}{\bibfnamefont{M.}~\bibnamefont{Wan}}, \bibnamefont{and}
  \bibinfo{author}{\bibfnamefont{Z.}~\bibnamefont{Xiao}},
  \bibinfo{journal}{Physical review letters} \textbf{\bibinfo{volume}{96}},
  \bibinfo{pages}{084502} (\bibinfo{year}{2006}).

\bibitem[{\citenamefont{Casciola et~al.}(2007)\citenamefont{Casciola,
  Gualtieri, Jacob, and Piva}}]{casciola2007residual}
\bibinfo{author}{\bibfnamefont{C.~M.} \bibnamefont{Casciola}},
  \bibinfo{author}{\bibfnamefont{P.}~\bibnamefont{Gualtieri}},
  \bibinfo{author}{\bibfnamefont{B.}~\bibnamefont{Jacob}}, \bibnamefont{and}
  \bibinfo{author}{\bibfnamefont{R.}~\bibnamefont{Piva}},
  \bibinfo{journal}{Physics of Fluids (1994-present)}
  \textbf{\bibinfo{volume}{19}}, \bibinfo{pages}{101704}
  (\bibinfo{year}{2007}).

\bibitem[{\citenamefont{Jacob et~al.}(2008)\citenamefont{Jacob, Casciola,
  Talamelli, and Alfredsson}}]{jacob2008scaling}
\bibinfo{author}{\bibfnamefont{B.}~\bibnamefont{Jacob}},
  \bibinfo{author}{\bibfnamefont{C.~M.} \bibnamefont{Casciola}},
  \bibinfo{author}{\bibfnamefont{A.}~\bibnamefont{Talamelli}},
  \bibnamefont{and} \bibinfo{author}{\bibfnamefont{P.~H.}
  \bibnamefont{Alfredsson}}, \bibinfo{journal}{Physics of Fluids
  (1994-present)} \textbf{\bibinfo{volume}{20}}, \bibinfo{pages}{045101}
  (\bibinfo{year}{2008}).

\bibitem[{\citenamefont{Frisch and Parisi}(1985)}]{Frisch1985}
\bibinfo{author}{\bibfnamefont{U.}~\bibnamefont{Frisch}} \bibnamefont{and}
  \bibinfo{author}{\bibfnamefont{G.}~\bibnamefont{Parisi}},
  \bibinfo{journal}{Turbulence and predictability in geophysical fluid dynamics
  and climate dynamics} pp. \bibinfo{pages}{84--88} (\bibinfo{year}{1985}).

\bibitem[{SI()}]{SI}
\emph{\bibinfo{title}{Supplemental information}}.

\bibitem[{\citenamefont{Meneveau and
  Sreenivasan}(1987)}]{meneveau1987multifractal}
\bibinfo{author}{\bibfnamefont{C.}~\bibnamefont{Meneveau}} \bibnamefont{and}
  \bibinfo{author}{\bibfnamefont{K.}~\bibnamefont{Sreenivasan}},
  \bibinfo{journal}{Nuclear Physics B-Proceedings Supplements}
  \textbf{\bibinfo{volume}{2}}, \bibinfo{pages}{49} (\bibinfo{year}{1987}).

\end{thebibliography}

\end{document}